\documentclass[12pt,preprint]{aastex}
\usepackage{epsfig}

\def\vkm{km s$^{-1}$}

\def\mJyb{mJy beam$^{-1}$}
\def\Jybk{Jy beam$^{-1}$ km s$^{-1}$}

\def\arcs#1{$#1''$}
\def\arcsa#1#2{$#1^{\prime\prime}_{^\textrm{.}}#2$}

\def\ra#1#2#3#4{$#1^\mathrm{h} #2^\mathrm{m} #3^\mathrm{s}_{^\textrm{.}} #4$}
\def\dec#1#2#3#4{$#1\degr #2\arcmin #3^{\prime\prime}_{^\textrm{.}}#4$}
  
\def\leftblank#1{}

\def\Ha{H$\alpha$}
\def\nHC3N{HC$_3$N}
\def\H13CCCN{H$^{13}$CCCN}
\def\HC13CCN{HC$^{13}$CCN}
\def\HCC13CN{HCC$^{13}$CN}

\def\CH2CN{CH$_2$CN}
\def\C2H3CN{CH$_2$CHCN}

\newcommand{\degree}{$^{\circ}$}

\def\VLSR{V_\textrm{\scriptsize LSR}}
\def\Vsys{V_\textrm{\scriptsize sys}}

\def\mH2{m_{\textrm{\scriptsize H}_2}}
\def\nH2{n_{\textrm{\scriptsize H}_2}}
\def\vkm{km s$^{-1}$}
\def\smpy{M$_\odot$ yr$^{-1}$}

\def\solarM{M_\odot}

\def\Voff{V_\textrm{\scriptsize off}}

\def\putfig#1#2#3{\epsfig{scale=#1,angle=#2,figure=#3}}

\begin{document}

\title{Multiple Fast Molecular Outflows in the PPN CRL 618}

\author{Chin-Fei Lee\altaffilmark{1},
Raghvendra Sahai\altaffilmark{2}, Carmen S\'anchez
Contreras\altaffilmark{3}, Po-Sheng Huang\altaffilmark{1}, and
Jeremy Jian Hao Tay\altaffilmark{4}}
\altaffiltext{1}{Academia Sinica Institute of Astronomy and Astrophysics,
P.O.  Box 23-141, Taipei 106, Taiwan}
\altaffiltext{2}{
Jet Propulsion Laboratory, MS 183-900, California Institute of Technology,
Pasadena, CA 91109, USA}
\altaffiltext{3}{
Astrobiology Center (CSIC-INTA), ESAC Campus, E-28691 Villanueva de la
Canada, Madrid, Spain
}
\altaffiltext{4}{
Department of Physics, National University of Singapore,
2 Science Drive 3, Singapore 117542
}

\begin{abstract}

CRL 618 is a well-studied pre-planetary nebula. It has multiple highly
collimated optical lobes, fast molecular outflows along the optical lobes,
and an extended molecular envelope that consists of a dense torus in the
equator and a tenuous round halo.  Here we present our observations of this
source in CO J=3-2 and HCN J=4-3 obtained with the Submillimeter Array at up
to $\sim$ \arcsa{0}{3} resolutions.  We spatially resolve the fast
molecular-outflow region previously detected in CO near the central star and
find it to be composed of multiple outflows that have similar dynamical ages,
and are oriented along the different optical lobes.  We also detect fast molecular
outflows further away from the central star near the tips of the extended
optical lobes and a pair of equatorial outflows inside the dense torus. 
We find that two episodes of bullet ejections in different directions
are needed, one producing the fast molecular outflows near the central star,
and one producing the fast molecular outflows near the tips of the extended
optical lobes.  One possibility to launch these bullets is the
magneto-rotational explosion of the stellar envelope.

\end{abstract}


\section{Introduction}

Pre-planetary nebulae (PPNs) are objects in the transition phase between the
asymptotic giant branch (AGB) phase and the planetary nebula (PN) phase in
the evolution of low- to intermediate-mass stars.  They are expected to
become planetary nebulae (PNs) in less than 1000 years as the central stars
become hot white dwarfs that can photoionize them.  Most PPNs and young PNs
were found to have highly aspherical shapes, with a significant fraction
having highly collimated bipolar or multipolar lobes
\citep{Corradi1995,Schwarz1997,ST98,Sahai01,Sahai2007}.  As a result,
instead of spherical fast winds as in the generalized interacting stellar
winds (GISW) model \cite[see e.g., review by][]{Balick2002}, collimated fast
winds (CFWs) have been proposed to operate during the post-AGB phase (and
even earlier during the late AGB phase), and be the primary agents for the
shaping of PPNs and young PNs \citep{ST98,Sahai01}.  Various CFW models have
been used to account for the morphology and kinematics of a few well-studied
PPNs and PNs.  These include a collimated radial wind with a small opening
angle \citep{Lee2003,Akashi2008,Lee2009}, a cylindrical jet
\citep{Cliffe95,Steffen98,Lee2004,Guerrero2008}, and a
bullet-like ejection \citep{Dennis2008}.  In this paper, we examine
these models with our high-resolution observations of the well-studied PPN
CRL 618, with the goal of understanding the mass-loss processes during the
end phases of the evolution of low and intermediate-mass stars.

CRL 618 is a nearby ($\sim$ 900 pc) well-studied PPN that has multiple
highly collimated optical lobes expanding rapidly away from the central star
\citep{Trammell2002,Sanchez2002}.  Infrared observations in H$_2$
\citep{Cox03} and (sub)mm-wave observations in CO (J=2-1, Sanchez et al.  2004,
hereafter SCetal04; J=6-5 and J=2-1, Nakashima et al.  2007) revealed fast
molecular outflows along the axes of the optical lobes, suggesting the
presence of underlying CFWs in this PPN.  The fast molecular outflows are
also prominent in high-J (J=16-15 and 10-9) CO transitions as shown in
Herschel observations \citep{Bujarrabal2010}.  MM-wave observations in CO
also showed low-velocity molecular cavity walls around the optical lobes
\citep{Sanchez2004}.  An extended molecular envelope has been detected
around the PPN, composing of AGB material ejected during the AGB phase.  It
can be divided into two structurally distinct components: (1) a dense torus
extending to $\sim$ \arcs{5} out from the central star in the equatorial
plane \citep{Sanchez2004b} and (2) a tenuous round halo surrounding the
torus \citep{Sanchez2004}.  A dense core is also seen near the central star
with an outer radius of $\sim$ \arcs{1} \citep{Sanchez2004,Lee2013}, tracing
the innermost part of the torus.  Recent STIS spectroscopic observations
showed that the optical forbidden line emission in the lobes is consistent
with shocks produced by the underlying CFWs \citep{Riera2011}.

Our hydrodynamical simulations showed that CFWs can readily produce the
morphology and the leading bow shocks of the optical lobes \citep{Lee2003}. 
In those simulations, the CFWs were assumed to be collimated radial winds
with a small opening angle.  However, in order to reproduce the observed
high-velocity optical emission inside the outflow lobes, the CFWs might need
to be more collimated, i.e., like cylindrical jets.  Alternatively,
the CFWs could also be bullet-like ejections as favored by
\citet{Dennis2008} in order to explain the multipolar morphology of this
PPN.

With the Submillimeter Array (SMA), we have observed this PPN in 350 GHz
band at unprecedented resolutions of up to $\sim$ \arcsa{0}{3}, providing
high-quality maps not only for many carbon-chain molecules such as HC$_3$N
and HCN and their isotopologues, but also for CO, HCO$^+$, and CS.  HC$_3$N
and their isotopologues trace mainly the dense core around the central star
and their results have been presented in \citet{Lee2013}.  Here we present
our maps of the CO J=3-2 and HCN J=4-3 lines, which have been found to be
the best probes of the fast molecular outflows in CRL 618.  Other lines such
as HCO$^+$ J=4-3 and CS J=16-15 trace roughly the same part of the outflows
as the CO and HCN lines, and thus will not be reported here.  With about 2
times higher resolution than that in \citet{Sanchez2004}, we spatially
resolve the fast molecular outflows previously mapped in CO near the central
star, with unprecedented precision. We also detect fast molecular outflows
further away from the central star at the ends of the extended optical lobes
and a pair of equatorial molecular outflows inside the dense torus.  We
derive the physical properties of the fast molecular outflows, and then
discuss the underlying CFWs and the implied mass-loss process.  The
equatorial outflows are first seen in this PPN and may have the same origin
as the fast molecular outflows.

\section{Observations}\label{sec:obs}


The observations toward CRL 618 were carried out on 2011 January 23 and
February 4 with the SMA in the very extended and extended configurations,
respectively.  The details of the observations have been reported in
\citet{Lee2013} and thus will not be repeated here.  Here we only report
important information for the CO J=3-2 and HCN J=4-3 maps to be presented in
this paper.  With natural weighting, we obtained a synthesized beam
(resolution) of size $\sim$ \arcsa{0}{53}$\times$\arcsa{0}{35} at a
position angle, PA$\sim$ 83\degree{}.  With robust weighting, we
obtained a synthesized beam of size $\sim$
\arcsa{0}{42}$\times$\arcsa{0}{26} at PA$\sim$ 82\degree{}.  The
channel maps are binned to have velocity resolutions of 1.41 \vkm{} and
16.91 \vkm{} per channel.  The rms noise levels are $\sim$ 45 \mJyb{} (or
2.54 K) in CO and 60 \mJyb{} (or 3.13 K) in HCN for the channel maps with a
resolution of 1.41 \vkm{} per channel.  The velocities of the channel maps
are LSR. In CRL 618, the systemic
velocity is $\Vsys = -21.5$ \vkm{} LSR, as assumed in
\citet{Lee2013}.  We also define an offset velocity $\Voff =
\VLSR - \Vsys$ to facilitate our presentation.
The flux uncertainty is estimated to be $\sim$ 20\%.
 The coordinates of the central star at the center of the maps are
$\alpha_{(2000)}$=\ra{04}{42}{53}{58},
$\delta_{(2000)}$=\dec{36}{06}{53}{38}.



\begin{figure} [!hbp]
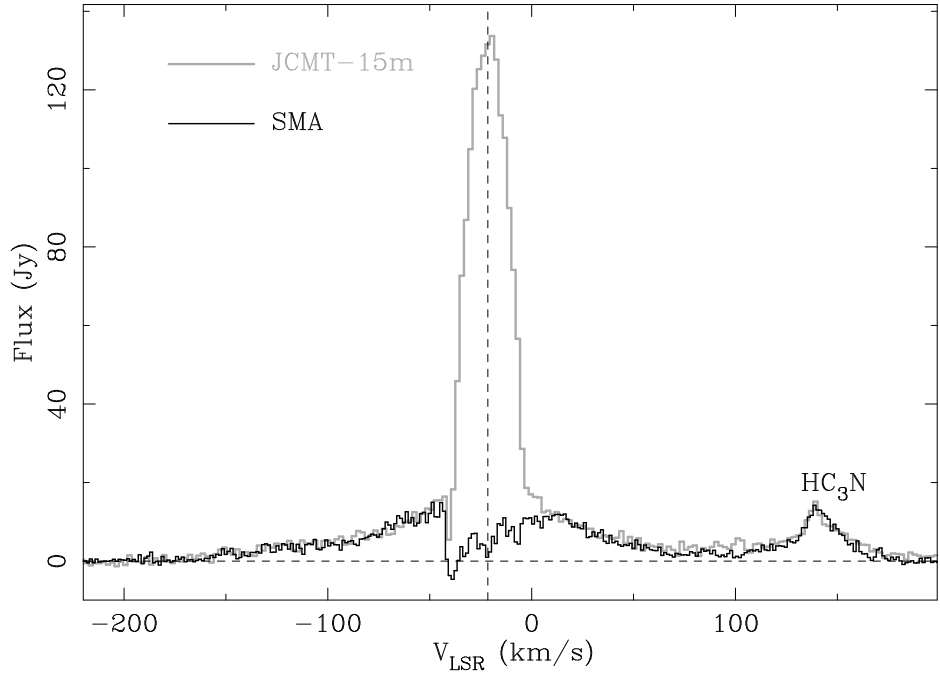

\centering
\putfig{0.5}{270}{f1.ps} 
\figcaption[]
{CO J=3-2 spectra obtained with the SMA (dark) and the JCMT 15m radio
telescope (gray, from JCMT archive) integrated over CRL 618. 
The vertical dashed line indicates the systemic velocity.
 In order to match the intensity
of the JCMT spectrum at high velocity, our SMA spectrum has been multiplied by a factor of
1.5.  The intensity peak at $\VLSR \sim$ 140 \vkm{} belongs to a \nHC3N{} line. 
\label{fig:COspec}
}
\end{figure}

Figure \ref{fig:COspec} shows our CO J=3-2 spectrum integrated over the
whole nebula of CRL 618 obtained with the SMA, overlaid on top of that
extracted from the JCMT 15m telescope data archive.  The JCMT spectrum was
an average of 5 spectra taken on 2009 November 29 and 2010 January 10.  The
flux intensity of our SMA spectrum has been multiplied by a factor of 1.5,
in order to match that of the JCMT spectrum at high velocity in the line
wings.  This indicates that at high velocity the flux in our SMA data is
two-thirds of that in the JCMT data.  Our SMA observations, with the
shortest baseline having a projected length of $\sim$ 45 m, are insensitive to
structure with a scale size of $\gtrsim$ \arcs{2}.  The flux loss
for the high-velocity emission, that has a size of $\lesssim$ \arcs{1} (see
next section), is insignificant.  That the line profile of the SMA spectrum
is almost the same as the JCMT spectrum in the line wings further supports
this argument.  As a result, the flux difference between the SMA data and
JCMT data is likely mainly due to the flux uncertainties in both our SMA
data ($\sim$ 20\%) and the JCMT data ($\sim$ 15\%), and some possible flux
lost in our SMA data.

The peak at $\VLSR \sim$ 140 \vkm{} traces \nHC3N{}
emission arising near the central star \citep{Lee2013}, spatially different
from the CO emission at that velocity and thus removed from
our CO maps.  At low velocity with $|\Voff| \lesssim$ 20 \vkm{},
the CO emission is mainly from the extended envelope
\citep{Sanchez2004} and thus its flux is mostly filtered out by the SMA.

\section{Observational Results}

In the following, our CO and HCN maps from the SMA observations (see, e.g.,
Figure \ref{fig:Ha_CO}) are overplotted on the optical \Ha{} image of CRL
618 in order to provide the reader with a global view of the nebular
structure and relative distribution of the molecular envelope and the
shock-excited optical lobes.  The \Ha{} image was taken by the Hubble Space
Telescope (HST) in August 2009 (Camera=WFC3/UVIS1, Filter=F656N, exposure
time=560s, pixel size=\arcsa{0}{04}), which was only $\sim$ 1.5 years
earlier than our SMA observations.  Therefore, the expansion of the optical
lobes at the tips in the \Ha{} image is only $\sim$ \arcsa{0}{1} \cite[a
mean proper motion of $\sim$ \arcsa{0}{5} in 7 years was obtained
by][]{Balick2013} and thus can be ignored in our study.  The \Ha{} image
shows precise geometry, structure, and orientation of the optical lobes in
the east (E1), southeast (E2), northeast (E3), northwest (W1), southwest
(W2), and other possible lobes E4 and W3 as well.  It also shows a clear
dust lane in the equatorial plane, tracing the dense torus there that has a
radius of $\sim$ \arcs{5} and a half thickness of $\sim$ \arcsa{1}{5}
\citep{Sanchez2004b}.  The optical lobes in the east are brighter and
visible closer to the central star than those in the west because the dusty
torus is tilted with its nearside toward the west.

\subsection{Molecular outflows}

Molecular outflows, which have been detected in CO in other transition lines
at J=2-1 and J=6-5 at lower angular resolution
\citep{Sanchez2004,Nakashima07}, are also detected here in the J=3-2 line at
higher angular resolution.  As seen in Figure \ref{fig:Ha_CO}a, CO emission
is mainly detected within $\sim$ $\pm$\arcs{2} from the central star,
tracing the outflows mainly inside the dense torus region, with blueshifted
emission to the east and redshifted emission to the west.  The emission
extending along the outflow axes of the E1, E2, E3, and W1 lobes reveals the
presence of fast molecular outflows associated with these lobes.  Redshifted
emission is also seen extending to the south in the equatorial plane and
will be discussed later with that seen in HCN.

\subsubsection{Molecular-Outflow Cavity Walls}


At low velocity (with $|\Voff| \lesssim $ 20 \vkm{}), the CO
map shows limb-brightened outflow shells that trace the outflow cavity walls
around the optical lobes, extending roughly equally to the east and west
from the central star to the dense torus (traced by the dust lane) and then
to the tenuous round halo where the optical lobes are seen (Fig. 
\ref{fig:Ha_CO}b).  Negative contours are seen in CO at low velocity partly
because extended emission is filtered out in our interferometric
observations and partly because of absorption by the expanding envelope. 
Some emission near the center could arise from the dense core detected in
the innermost part of the torus \citep{Lee2013}.  The walls are bright near
the star inside the dense torus, but becomes faint further out inside the
tenuous round halo.  This indicates that the wall brightness depends on the
density of the envelope, in agreement with that the walls are composed
mainly of the envelope material swept up by underlying CFWs, as suggested in
\citet{Sanchez2004}.  The walls appear less extended than those seen in CO
J=2-1 \cite[see Fig.  8 in][]{Sanchez2004}, likely because the more extended
part of the walls is more tenuous and cooler, and thus better traced in
lower transition line.


\subsubsection{Fast Molecular Outflows} \label{sec:fmo}

A pair of fast molecular outflows have been detected previously in CO near
the central star, one in the east and one in the west
\citep{Sanchez2004,Nakashima07}.  Their structure and kinematics can be
seen in our CO channel maps presented in Figures \ref{fig:COchan_big} and
\ref{fig:COchan} (for zoom-in to the central region). The fast molecular
outflow in the west is now spatially resolved also in the N-S
direction.  The emission with $\Voff\gtrsim$ 120 \vkm{} forms a bow-like
structure projected inside the W1 lobe, with a tip at $\sim$ \arcsa{2}{3} to
the west pointing away from the central star (see Fig.  \ref{fig:COchan} and
also Fig.  \ref{fig:Ha_CO}d).  The velocity decreases from the bow tip to
the bow wings toward the central star.  On the other hand, the emission with
$\Voff\lesssim$ 100 \vkm{} (see the channel maps from 70.4 to 2.81 \vkm{} in
Fig.  \ref{fig:COchan}, and also yellow contours in Fig.  \ref{fig:Ha_CO}d)
forms two limb-brightened shells upstream of the bow-like structure.  The
shells are roughly aligned with the two limb-brightened edges of the W1
optical lobe further away from the central star.  They have an extent of
$\sim$ \arcsa{1}{5} to the west, similar to the dense torus, and thus are
likely composed mainly of the material in the dense torus.  As a result, the
shells likely trace cavity walls of the W1 lobe near the central star,
produced by the bow-like structure.
 Note that the southern shell lies on the W2 lobe and thus part of it could
trace cavity walls of the W2 lobe as well.

The fast molecular outflow in the east is now resolved into three fast
molecular outflows, one inside the E1 lobe, one inside the E2 lobe, and one
on the E3 lobe (see also Fig.  \ref{fig:Ha_CO}c).  The one inside the E1
lobe and the one on the E3 lobe overlap with each other near the central
star.  The three outflows are compact with their tips at a distance of
\arcsa{2}{2} to \arcsa{2}{5} from the central star.  Their velocity also
decreases toward the central star (see also Figs.  \ref{fig:Ha_CO}c and
\ref{fig:Ha_CO}d).  The fast molecular outflow on the E3 lobe is slightly
resolved, appearing as a collimated lobe that closes back to the central
star (Fig.  \ref{fig:Ha_CO}d), surrounding the E3 optical lobe.  Cavity
walls are also seen in the east within $\sim$ \arcsa{1}{5} from the central
star surrounding the E1/E2/E3 lobes (see the channel maps from $-$116 to
$-$47.9 \vkm{} in Fig.  \ref{fig:COchan}).

In addition, fast molecular outflows are also detected further away from the
central star at the end of the E1 and W2 lobes (see Figs. 
\ref{fig:COchan_big} and \ref{fig:Ha_CO}c).  They are faint in CO J=3-2
emission and thus not detected before in other interferometric studies. 
They are collimated and elongated, lying inside the optical lobes.


We now investigate in detail, the spatio-kinematic structure of the fast 
molecular outflows using
position-velocity (PV) diagrams. We first examine cuts in our CO J=3-2 maps, 
taken along the axes of optical lobes shown in 
Figure \ref{fig:Ha_CO}. In Figure
\ref{fig:pvs_jet}a, which shows cuts along the axes of the E1 and W1 optical 
lobes, we find two prominent linear 
PV structures (delineated by
two thick solid lines) extending from position offsets of 
$\pm$\arcsa{1}{5} to $\pm$\arcsa{2}{3}. These are associated with the fast
molecular outflows in the E1/E3 and W1 lobes near the central star.  
A third linear PV structure (solid line) seen at position offsets in the range  
\arcsa{4}{0} to \arcsa{5}{7}, is associated
with the fast molecular outflow near the end of the E1 lobe. In Figure
\ref{fig:pvs_jet}b, which shows cuts 
along the axes of the E2 and W2 optical lobes, we see 
two linear PV structures (solid lines) with position offsets in the ranges  
\arcsa{1}{7} to \arcsa{2}{7}, and -\arcsa{3}{0} to 
-\arcsa{4}{6}, that are associated
with the fast molecular outflows in these lobes.  Therefore, all of the
fast molecular outflows are associated with a linear PV structure,
indicating that their velocity decreases roughly linearly from their tips
towards the central star. The PV structures with position offsets $\lesssim \pm$\arcsa{1}{5} 
from the central star
likely trace the cavity walls near the central star.  


Next, we examine cuts taken laterally across the lobes at 
different radial distances from the central 
star (Fig. \ref{fig:pvs_perjet}). The right panels in this figure show
the PV diagrams for the fast molecular outflow in the W1 lobe. In the cut  
near the tip of the lobe (panel $p$),
the emission peak occurs at the high-velocity end of the PV structure. But
in the cuts within $\sim$ \arcsa{1}{6} of the central star (panels $m$ to
$h$) , the emission peak shifts towards the
low-velocity end of the PV structure.  This is because the emission
at a radial distance of $\lesssim$ \arcsa{1}{6} mainly traces 
material in the cavity walls;  panels  $m$ to
$h$ reveal an asymmetric hollow PV structure with its cross-section
first increasing from the systemic velocity to $\sim$ $-$10 \vkm{}, and
then decreasing towards the high redshifted velocity. 

The cuts across the fast
molecular outflow in the E1 lobe (Fig.  \ref{fig:pvs_perjet}, left) 
show a similar PV structure, except that 
it occurs at blueshifted velocities. As in the case of the W1 lobe, the 
emission in the E1 at a radial distance $\lesssim$\arcsa{1}{6} from the central star, comes 
mainly from the cavity walls. We note that the fast molecular outflow along the 
E2 (E3) lobe is also seen in panel $g$ (panels $e$ and $f$).


\subsection{Comparison with previous interferometric results}

In previous CO interferometric observations \cite[e.g,][]
{Sanchez2004,Nakashima07}, the fast molecular outflows were seen as two
blobs near the central star, one in the west and one in the east, and thus
considered as a filled high-velocity bipolar outflow.  In our observations
with unprecedented resolutions, the fast molecular outflows are better
resolved in the N-S direction, allowing us to compare the CO outflow
structure with the HST H$\alpha$ image in more detail.  The one in the west
now appears as a bow-like structure projected inside the W1 lobe.  The one
in the east is resolved into three outflows, one inside the E1 lobe, one
inside the E2 lobe, and one on the E3 lobe.  In addition, judging from the
detailed structure and kinematics of the fast molecular outflows, we argue
that the fast molecular outflows near the star with a distance smaller than
$\sim$ \arcsa{1}{6} (or equivalently with $|\Voff|\lesssim$ 120
\vkm{}) are more likely to trace the cavity walls composed mainly of the
material in the dense torus.  Moreover, at large distance from the central
star, we detect two more fast molecular outflows, one near the end of the E1
lobe and one near the end of the W2 lobe.

In summary, our study provides a very important qualitative improvement over
previous one by showing that individual fast molecular outflows are
associated with individual optical lobes.  Thus, our study shows that both
the optical and mm-wave observations are tracing the same physical
structures.  In contrast, the previous studies proposed a model that
consists of a single bipolar fast molecular outflow inside a single bipolar
elliptical cavity.

\subsection{A Pair of Equatorial Clumps}




Molecular outflows are also detected in HCN J=4-3, as shown in Fig. 
\ref{fig:eqoutflow}a, similar to that seen in CO.  At low velocity, similar
molecular cavity walls are also seen around the optical lobes.  In addition,
a pair of clumps are also seen in the equatorial plane, with the
blueshifted emission in the north and the redshifted emission in the south
(Fig.  \ref{fig:eqoutflow}b).  They are spatially unresolved perpendicular
to the equatorial plane.  They are bright with a peak brightness temperature
of $\sim$ 50 K.  They are also detected in CO (Fig.  \ref{fig:Ha_CO}b), but
the northern clump is almost missing, likely because of substantial
absorption by the cold extended expanding envelope seen in CO
\citep{Sanchez2004}.  The emission in the southern clump is not absorbed by
the extended envelope because the latter does not have foreground material
at the radial velocity of the former.

The kinematics of these clumps can be studied using PV diagrams from cuts
taken along the equatorial plane at P.A.=0\degree{} in HCN and CO (Fig. 
\ref{fig:pvs_equatorial}).  The PV diagrams for HCN and CO are similar.  The
PV structures in the dashed boxes in the figure are associated with the two
clumps.  The same foreground absorption that attenuates the CO emission from
the northern clump (due to the cold envelope), produces an emission gap at
$\sim$ $-$33 \vkm{} in the HCN map.  The emission from the two clumps
extends from near the systemic velocity to $\sim$ $\pm 25$ \vkm{} away.  The
redshifted emission in the south clump extends from a radial distance of
\arcsa{0}{8}--\arcsa{1}{6} and the blueshifted emission in the north clump
extends \arcs{1}--\arcs{2}.  No clear velocity gradient is seen within each
clump along the equatorial plane.

\section{Discussion}

\subsection{Fast molecular outflows: Total mass}



Fast molecular outflows are seen around and inside the optical lobes.  Their
mass can be estimated from their CO emission.  Assuming that the CO emission
arises from LTE gas, then the mass is given by
\begin{eqnarray} 
m &=&
\frac{m_{\textrm{\scriptsize H}_2}}{X} \frac{d^2}{I_J(\textrm{CO})}
\frac{\nu}{c}(10^{-18} F) c_\tau \eta
\end{eqnarray} 
where $F$ is the CO flux density in Jy \vkm{},
$d$ is the distance, $X$ is the CO-to-H$_2$ relative abundance of
$2\times10^{-4}$ as adopted in \citet{Sanchez2004}, $c_\tau$ is the
optical depth correction factor defined as
\begin{equation} 
c_\tau = \frac{\tau}{1-e^{-\tau}} 
\end{equation} 
and $\eta=1.5$ to account for the flux scale uncertainty and possible flux
lost in our SMA data as mentioned earlier in Section \ref{sec:obs}. The
emission coefficient of CO J=3-2 line is
\begin{equation}
I_J (\textrm{CO})= \frac{h \nu}{4 \pi} \frac{g_J \exp (-E_u/kT)}{Q} A_{32}
\end{equation}
where $Q$ is the partition function. Assuming that the gas has a temperature
of $T=200$ K \citep{Bujarrabal2010}, then $I_J(\textrm{CO}) =
3.82\times10^{-23}$ erg s$^{-1}$ sr$^{-1}$ mol$^{-1}$.

The optical depth of the CO emission can be estimated from the
single-dish flux ratio of CO J=3-2 (as shown in Figure \ref{fig:COspec}) to
CO J=2-1 (as shown in Figure 1 in \citet{Sanchez2004}) in the line wings
with $|\Voff| > 20$ \vkm{}.  The flux ratio
 there is found to be $\sim$ 3.  At $T=200$ K, this requires the optical
depth of CO J=3-2 to be $\sim$ 2.5.  Thus, we have $c_\tau \sim 2.7$.  
As a result, the mass can be simply given by
\begin{eqnarray} 
m &\approx& 7.86\times10^{-5} F \;\; (\solarM) 
\label{eq:fmomass}
\end{eqnarray} 


Here we estimate the mass for the fast molecular outflows that have an
expansion velocity greater than 20 \vkm{}.  The fluxes of the fast molecular
outflows in the east (E1, E2 and E3 lobes) and west (W1 and W2 lobes) are
435.6 and 391.7 Jy \vkm{}, respectively.  Thus, the total flux is 827.3 Jy
\vkm{}, giving a total mass of 6.5$\times10^{-2}$ $\solarM$, slightly lower
than the value (0.09 $\solarM$) derived by \citet{Sanchez2004} integrating
the CO profile over the whole width of the emission wings, i.e.,  over a
velocity range larger than that used by us.


\subsubsection{CFWs} \label{sec:CFWs}



The fast molecular outflows can be used to probe the underlying CFWs that
produce the optical lobes.  In current simulations, the CFW can be a
collimated radial wind with a small opening angle \citep{Lee2003,Lee2009}, a
cylindrical jet with a velocity parallel to the jet axis \citep{Lee2001}, or
a bullet (a massive clump) \citep{Dennis2008}.  In each of these three
cases, we expect a bow shock to be formed at the head of the CFW, producing
a collimated lobe composed of the bow shock at the tip and cavity walls at
the base, with the velocity decreasing from the tip to the base.  


Such expectation is supported by the fast molecular outflow in the E3 lobe,
that has a collimated shape closing back to the central star, with the
velocity decreasing from the tip towards the central star.  The bow shock in
this lobe can be traced by the highest velocity material seen at the tip of
the lobe (Figure \ref{fig:Ha_CO}d).  The fast molecular outflow seen in
projection on the W1 lobe is likely physically inside the latter.  In this
case, its bow-like structure can be an internal bow shock produced either by
a temporal variation in the wind/jet velocity
\citep{Lee2001,Lee2003,Lee2009} or by an internal bullet
\citep{Lee2004,Yirak2009}.  The wings of this internal bow shock can
interact with (and perhaps brighten) the outflow cavity walls upstream.  
The PV plots from cuts in our CO J=3-2 images taken across the cavity walls
show an asymmetric hollow structure (see Sec.  \ref{sec:fmo}), as expected if the
cavity walls are produced by a CFW.
The tip of the fast molecular outflow near the central star in the E1 lobe may
be the signature of an internal bow shock as well.  Other fast molecular
outflows that appear as collimated clumps or elongated structures inside the
optical lobes may also trace internal bow shocks, but are spatially
unresolved in our observations. 


\subsubsection{Dynamical Age}



Table \ref{tab:mass_age} lists the dynamical ages of the fast molecular
outflows that can trace the (internal) bow shocks, estimated using the
tip radial velocity ($V_t$) and tip distance ($P_t$) measured from the
PV diagrams, corrected for the inclination, with the following formula
\begin{equation}
t_\textrm{\scriptsize dyn} = \frac{P_t}{V_t} \tan i
\end{equation}
where $i$ is the inclination angle to the plane of the sky.


We set constraints on the inclination angles for individual lobes using
previously published results.  By modelling the spatio-kinematic structure
seen in the bipolar high-velocity outflow HCO$^+$ J=1-0
emission\footnote{this study had relatively low angular resolution and could
not separate out the contributions from individual lobes on the eastern or
western sides of the center}, \citet{Sanchez2004b} derive an average
inclination of 32\degree{}.  From long-slit optical spectroscopy of the
lobes, \citet{ Sanchez2002} estimate that the extended optical lobes, E1 and
E2, have inclination angles of 24\degree{}$\pm$6\degree{}.  The extended
optical lobes, W1 and W2, appear to be the counterparts of the E2 and E1
lobes on the west side of the nebula, hence we assume they have similar
inclination angles.
As a result, we assume an inclination
angle of 30\degree{}$\pm$10\degree{} for the fast molecular outflows inside
the extended optical lobes E1, E2, W1, and W2.  
Compared to the E1, E2, W1, W2 lobes, the E3 lobe appears to be much
shorter.  This could either be a result of projection - i.e., E3 could be
comparable in length to the other lobes and highly inclined, or it could be
intrinsically short (e.g., if it resulted from the same physical event that
created the inner outflows in E1, E2, and W1).  Thus, the inclination angle
for this lobe (and thus its fast molecular outflow)
is the most uncertain, and we assume it to be 30\degree{}$\pm$20\degree{}


The age estimates in Table \ref{tab:mass_age} fall into two groups, one with
a mean of 45$\pm$25 years, and another with a mean of 105$\pm$40 yrs.  We
therefore infer that there have been two episodes of CFWs in CRL618, a
recent one that occurred about 45 yrs ago producing the fast molecular
outflows near the central star, and an older one that occurred 105 yrs ago
producing the fast molecular outflows near the tips of the extended optical
lobes.

Even though there are uncertainties associated with our age estimates, we
think that the above inference is robust.  First, for the E1 lobe, the
difference between the ages of the inner and outer outflows (34 versus 109
years) cannot be reconciled due to inclination uncertainties because each of
these ages are affected in the same way by the common inclination factor. 
Second, our age estimates for the molecular outflows at the tips of the
extended optical lobes are in good agreement with that derived by
\citet{Balick2013} for the tips from their proper motion study.  They
derived a mean expansion (i.e, dynamical) age of 100$\pm$15 yrs, which does
not require a knowledge of the inclination.



\subsubsection{Mass and Mass-loss rate of CFWs}


Table \ref{tab:mass_age} also lists the masses of the fast molecular
outflows that can trace the (internal) bow shocks, derived using Eq. 
\ref{eq:fmomass}.  For the fast molecular outflows near the central star,
only their tips at a radial distance of more than \arcsa{1}{6} (or
correspondingly with $|\Voff| \gtrsim$ 120 \vkm{}, see Sec.  \ref{sec:fmo})
trace and thus are used to derive the masses of the (internal) bow shocks. 
As discussed in Sec.  \ref{sec:fmo}, the emission of the fast molecular
outflows within $\sim$ \arcsa{1}{6} of the central star likely traces the
cavity walls, and thus are not included in this mass estimate.  For other
fast molecular outflows that appear as collimated clumps or elongated
structures inside the optical lobes, all their emission can trace and be
used to derive the masses of the (internal) bow shocks.  The masses of the
molecular material in the (internal) bow shocks are on the order of
$10^{-3}$ $\solarM$.  The total mass of the internal bow shocks is
$\sim 7.5\times10^{-3}$ $\solarM$, which is about a tenth of the total
outflow mass.

We can now estimate the mass-loss rate in the CFWs, assuming that the
CFWs are mainly molecular as proposed by \citet{Lee2009}, and are seen as
the (internal) bow shocks.  As discussed earlier, there are two episodes of
CFWs.  For the recent one, the (internal) bow shocks near the central star
are used.  Their mean mass is $\sim$ 1.6$\times10^{-3}$ $\solarM$.  Their
duration is $\sim$ 15 yrs, estimated from their length ($\sim$ \arcsa{0}{8})
and their tip radial velocity, corrected for the inclination.  Thus, the
mass-loss rate is $\sim$ 1.1$\times10^{-4}$ \smpy.  Note that this mass-loss
rate is one order of magnitude lower than the one ($\gtrsim 10^{-3}$
\smpy) derived in \citet{Sanchez2004}, which used the total mass of the fast
molecular outflows.  For the older one, the (internal) bow shocks near the
end of the optical lobes are used.  Their mean mass is $\sim$
0.6$\times10^{-3}$ $\solarM$.  Their duration is $\sim$ 30 yrs, estimated
from their length ($\sim$ \arcsa{1}{5}) and tip radial velocity, corrected
for the inclination.  Thus, the mass-loss rate is $\sim$ 2$\times10^{-5}$
\smpy.  Note that our derived mass-loss rates are highly uncertain, because
the mass may include some envelope material, the excitation temperature is
uncertain, and molecular fraction is uncertain.  Nonetheless, the mass-loss
rate of the older fast molecular outflows turns out to be similar to that
recently used by \citet{Balick2013} in their jet and bullet models to
produce their CRL 618 observations of the extended optical lobes.


\subsubsection{Possible Origin}





In our first simulation study, we found that a collimated radial wind can
produce a highly collimated lobe similar to the W1 lobe, however, it has
difficulty producing the bright emission structures seen along the body of
the lobe \citep{Lee2003}.  In our second such study, we also found that a
collimated radial wind has difficulty producing the fast molecular outflow
inside the lobe near the central star \citep{Lee2009}.  As suggested in
those studies, a cylindrical jet may be needed to produce the W1 lobe.  

Now in our new observations, fast molecular outflows are detected at the
ends of the extended optical lobes, providing an additional constraint to
the shaping mechanism of this PPN.  They are found to have a similar
dynamical age of $\sim$ 105 yrs.  At about the same time as our study,
\citet{Balick2013} also find that the extended optical lobes all have a
similar dynamical age of $\sim$ 100 yr, similar to that of the fast
molecular outflows.  Thus, in order for the jet model to work, a set of jets
in different orientations are needed to be launched around the same time
(i.e, $\sim$ 100 yrs ago).  However, there is no known physical mechanism
that can launch such jets.  We therefore think that the bullet model that
has multiple bullets ejected simultaneously in different directions around
the polar (i.e., rotational) axis is needed, as favored by
\citet{Dennis2008} and \citet{Balick2013}.  One possibility to launch these
bullets is the magneto-rotational (MR) explosion of the stellar envelope
along the rotational axis \citep{Matt2006}, at the end of the AGB phase.



The fast molecular outflows near the central star in different directions
are also found to have a similar dynamical age of $\sim$ 45 yrs.  Thus,
another episode of bullet ejection seems to be needed.  However, it is
unclear if there could be a second episode of MR explosion.  Alternatively,
this recent bullet episode could be related to the onset of the H II region
or PN phase that was found to take place around the same time at $\sim$ 40
yrs ago \citep{Tafoya2013}.


\subsection{A Pair of Equatorial Outflows}







A pair of molecular clumps are seen in HCN and CO in the equatorial plane at
a radius of $\sim$ \arcsa{1}{5} (1350 AU) from the central star, with the
redshifted emission to the south and blueshifted emission to the north. 
They lie inside the extended, expanding dense torus detected in the J=1-0
lines of HCN and HCO$^+$ \citep{Sanchez2004b}. 
Their deprojected velocity can not be determined
because their inclination angle is unknown.  Their structure is spatially
unresolved perpendicular to the equatorial plane.


The two clumps have maximum  offset velocities of $\Voff = \pm$25
\vkm{}, which is too big to arise from rotation around the central star at
a radius of $\sim$ 1350 AU.  Therefore, they must trace a pair of collimated
molecular outflows (i.e., a collimated bipolar outflow) inside the expanding
dense torus.  Although it is possible that these clumps are simply
compact condensations in the halo not filtered out by the interferometry,
their diametrically opposed symmetric location around the nebular center
argues otherwise.  
Furthermore, 2.12 \micron{} H$_2$ line (S(1), v=1-0) emission images of CRL
618 by \citet{Cox03} show a clump at the same location as the northern
HCN clump, indicating the presence of shock-excited gas at that location. 
This supports our outflow hypothesis for these clumps, in which these are composed of
shocked, compressed material at the locations where the fast outflows have
been substantially decelerated by the slowly expanding, ambient
circumstellar envelope.  The lack of an H$_2$ emission counterpart to the
southern (red-shifted) HCN clump is most likely due to higher foreground
extinction by circumstellar dust, compared to the northern one (which is
blue-shifted and closer to us).  Similar clumpy outflows have also been seen
in the Egg Nebula (in the 2.12 \micron{} H$_2$ line and CO J=2-1 line) in its
dense, dusty equatorial torus \citep{Sahai1998,Cox2000,Balick2012b}.


These two clumps have a mass of $>$ 1.1$\times10^{-4}$ $\solarM$, assuming
that their HCN emission is optically thin with an excitation temperature of
100 K, and a HCN-to-H$_2$ relative abundance of $>2\times10^{-7}$ as derived
in \citet{Sanchez2004b}.  The MR explosion that we discussed earlier for the
fast molecular outflows produces an explosion simultaneously in the equator
as well \citep{Matt2006}.  However, that explosion may produce a torus-like
outflow instead of a bipolar outflow as seen here in the equator.



\section{Conclusions}


We have presented our SMA results of the molecular outflows of CRL 618 in CO
J=3-2 and HCN J=4-3 in the 350 GHz band at up to $\sim$ \arcsa{0}{3}
resolution.  With about 2 times higher resolution than that before, we have
better resolved the molecular outflow cavity walls and the fast molecular
outflows in this PPN, showing that individual fast molecular outflows are
associated with individual optical lobes.  We have detected a pair of
equatorial molecular outflows inside the dense torus, for the first time in
this PPN.  At high velocity, the fast molecular outflows could trace the
(internal) bow shocks and thus the collimated fast winds ejected from the
central star.  The multipolar morphology and the similar dynamical ages of
the fast molecular outflows together suggest the underlying collimated fast
winds to be massive bullets ejected from the central star.  There seems to
be two episodes of bullet ejections, a recent one at $\sim$ 45 yrs and an
older one at $\sim$ 105 yrs ago, producing the fast molecular outflows near
the central star and near the tips of the extended optical lobes,
respectively.  One possibility to launch these bullets is the
magneto-rotational explosion of the stellar envelope.


\acknowledgements 
We thank the anonymous referee for valuable suggestions.
We thank the SMA staff for their efforts in running and maintaining the
array.  C.-F.  Lee and P.-S.  Huang acknowledge grants from the National
Science Council of Taiwan (NSC 99-2112-M-001-007-MY2 and NSC
101-2119-M-001-002-MY3) and the Academia Sinica (Career Development Award). 
RS's contribution to the research described here was carried out at the Jet
Propulsion Laboratory, California Institute of Technology, under a contract
with NASA.  CSC has been partially supported by the Spanish MICINN/MINECO
through grants AYA2009-07304, AYA2012-32032, and CONSOLIDER INGENIO 2010 for
the team ``Molecular Astrophysics: The Herschel and ALMA Era -- ASTROMOL''
(ref.: CDS2009-00038).

\newpage 
\begin{deluxetable}{lcccccc}
\rotate
\tablecolumns{7}
\tabletypesize{\normalsize}
\tablecaption{Dynamical age, Flux, and Mass of fast molecular outflows that trace
(internal) bow shocks
\label{tab:mass_age}}
\tablewidth{0pt}
\tablehead{
\colhead{Lobe} 
 & \colhead{Tip Distance} & \colhead{Tip Radial Velocity} & \colhead{Dynamical Age}
 & $V_\textrm{\scriptsize LSR}$ Range & \colhead{Flux Density} & \colhead{Mass} \\
\colhead{} 
 & \colhead{(arcsec)} & \colhead{(\vkm{})} & \colhead{(yr)}
 & (\vkm) & \colhead{(Jy \vkm{})} & \colhead{($10^{-3}$ $\solarM$)}
}
\startdata
 E1       & \arcsa{2}{3} & 165 & 34$\pm13$  & $< -141.5$ & 22 & 1.74 \\ 
 E1 outer & \arcsa{5}{5} & 125 & 109$\pm43$ & $-$153.6 to $-$104.3 & 9.2 & 0.72\\ 
 E2       & \arcsa{2}{5} & 110 & 56$\pm22$  & $-$148.0 to $-$101.5 & 9.3 & 0.73\\ 
 E3       & \arcsa{2}{2} & 140 & 39$\pm27$  & $< -141.5$ & 18 & 1.41 \\
 W1       & \arcsa{2}{4} & 165 & 36$\pm14$  & $> 98.5$   & 31 & 2.44\\
 W2       & \arcsa{4}{5} & 110 & 101$\pm40$ & 69.0 to 90.2 & 5.7 & 0.45 
\enddata
\end{deluxetable}


\newpage

\begin{figure} [!hbp]
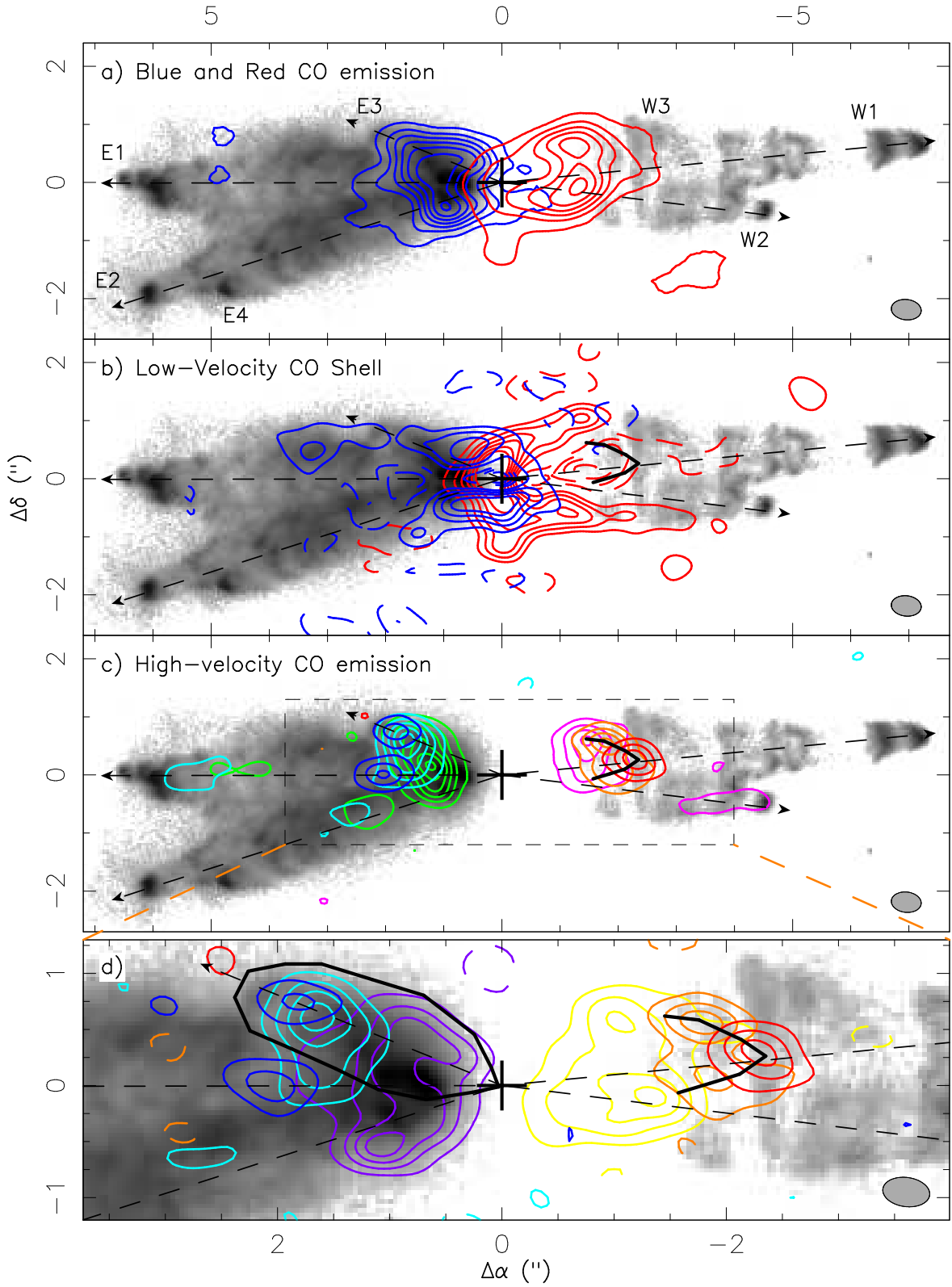

\centering
\putfig{0.9}{0}{f2.ps} 
\figcaption[]
{See next page.
}
\end{figure}

\setcounter{figure}{1}


\begin{figure} [!hbp]
\centering
\figcaption[]
{Our CO (contours) maps plotted on the \Ha{} image (gray in logarithmic
scale) that shows the optical lobes.  The cross marks the central star
position.  The dashed arrows indicate the outflow axes of the optical lobes. 
{\bf(a)} shows the maps of the redshifted (-21.5 to 157.8 \vkm) and blueshifted
(-194.4 to -21.5 \vkm) CO emission.  
{\bf(b)} shows the redshifted (-16.9 to -8.5 \vkm) and
blueshifted (-35.2 to -26.8 \vkm) CO outflow shell at low velocity. 
{\bf(c)} shows the 3
high-redshifted [R1 (red) to R2 (orange) to R3 (magenta)] and 3
high-blueshifted [B1 (blue) to B2 (cyan) to B3 (green)] CO emission
structures.  
{\bf(d)} shows the blow-up of the outflows in the central region at higher
angular resolution.  In order to show the bases of the W1 and E3 outflows,
we plot the redshifted emission in R4 (yellow) instead of R3, and
blueshifted emission in B4 (purple) instead of B3.
As indicated in Figure
\ref{fig:pvs_jet}, the velocity ranges are: (B1) -183.2 to -160.6, (B2) -157.8 to
-132.5, (B3) -132.5 to 114.1, (B4) -84.4 to -70.5, (R1) 128.2 to 150.8, (R2)
94.4 to 119.8, (R3) 70.4 to 88.8, and
(R4) 28.2 to 42.3 \vkm.  
The bow depicts the bow-like structure for the fast
molecular outflow in the W1 lobe, by connecting the peaks of the CO emission
with $\Voff > 120$ \vkm{} as seen in Figure \ref{fig:COchan}. 
The black lobe in {\bf(d)} circles the fast molecular outflow along the E3 lobe.
The angular resolution is
\arcsa{0}{52}$\times$\arcsa{0}{35} in {\bf (a)-(c)} and
\arcsa{0}{42}$\times$\arcsa{0}{26} in {\bf (d)}.  
The contours start at 4.2, 0.4, 1, and 0.7 \Jybk{} with a step of
8.4, 0.8, 2, and 1.4 \Jybk{}, respectively, in panel (a), (b), (c), and (d). 
Equivalent negative contours are plotted in dashed lines.
\label{fig:Ha_CO} }
\end{figure}



\begin{figure} [!hbp]
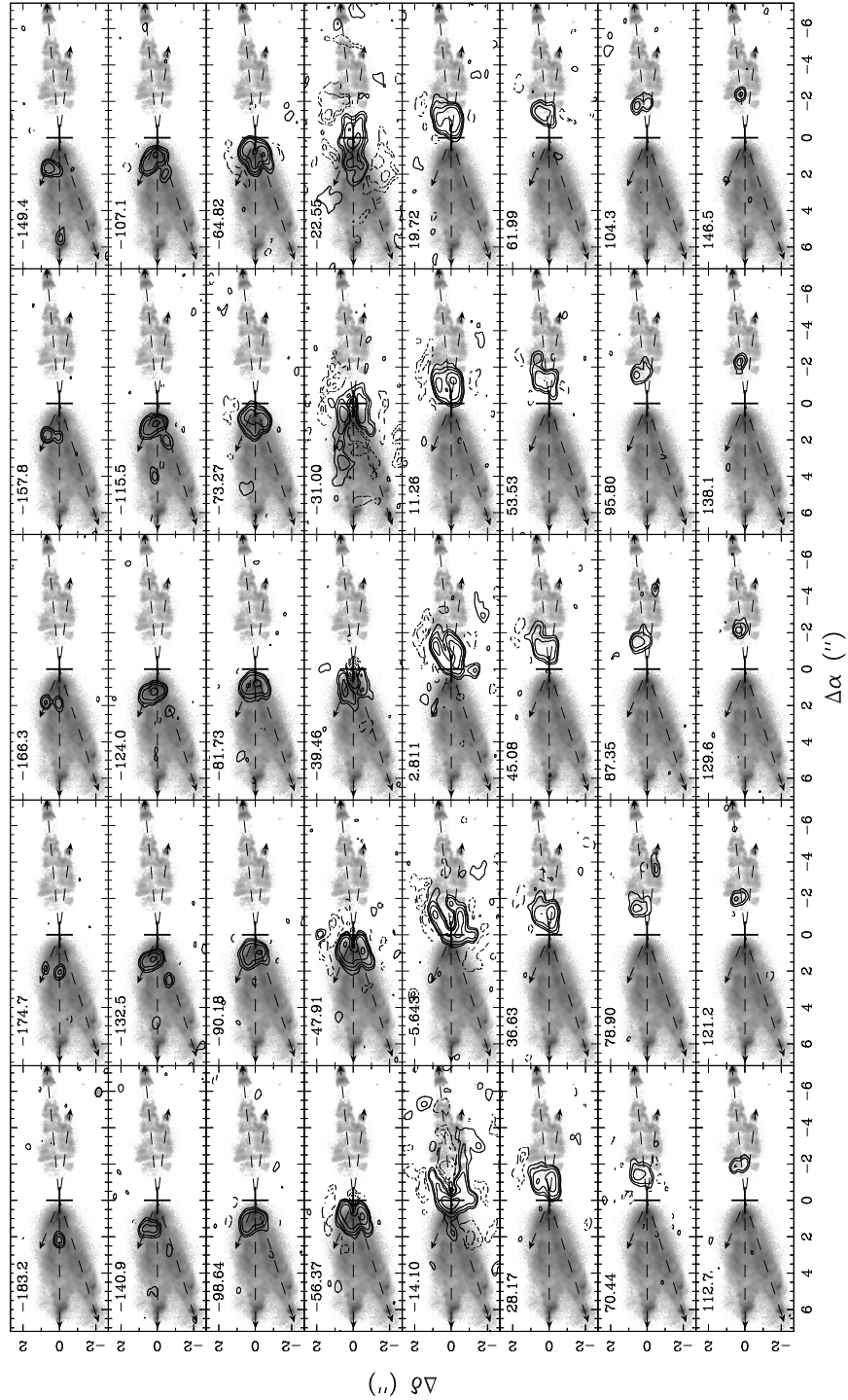

\centering
\putfig{0.8}{0}{f3.ps} 
\figcaption[]
{
CO channel maps on top of the H$\alpha$ image toward CRL 618.
The cross and arrows have the
same meaning as those in Figure \ref{fig:Ha_CO}.  The angular resolution is
\arcsa{0}{52}$\times$\arcsa{0}{35}. The velocity in \vkm{} is indicated in the
top left corner in each panel. The contour levels are
$3\cdot2^{n-1}\sigma$, where $n=1,2,3..$, and $\sigma=21$ \mJyb{}.
Equivalent negative contours are plotted in dashed lines.
\label{fig:COchan_big}
} 
\end{figure}

\begin{figure} [!hbp]
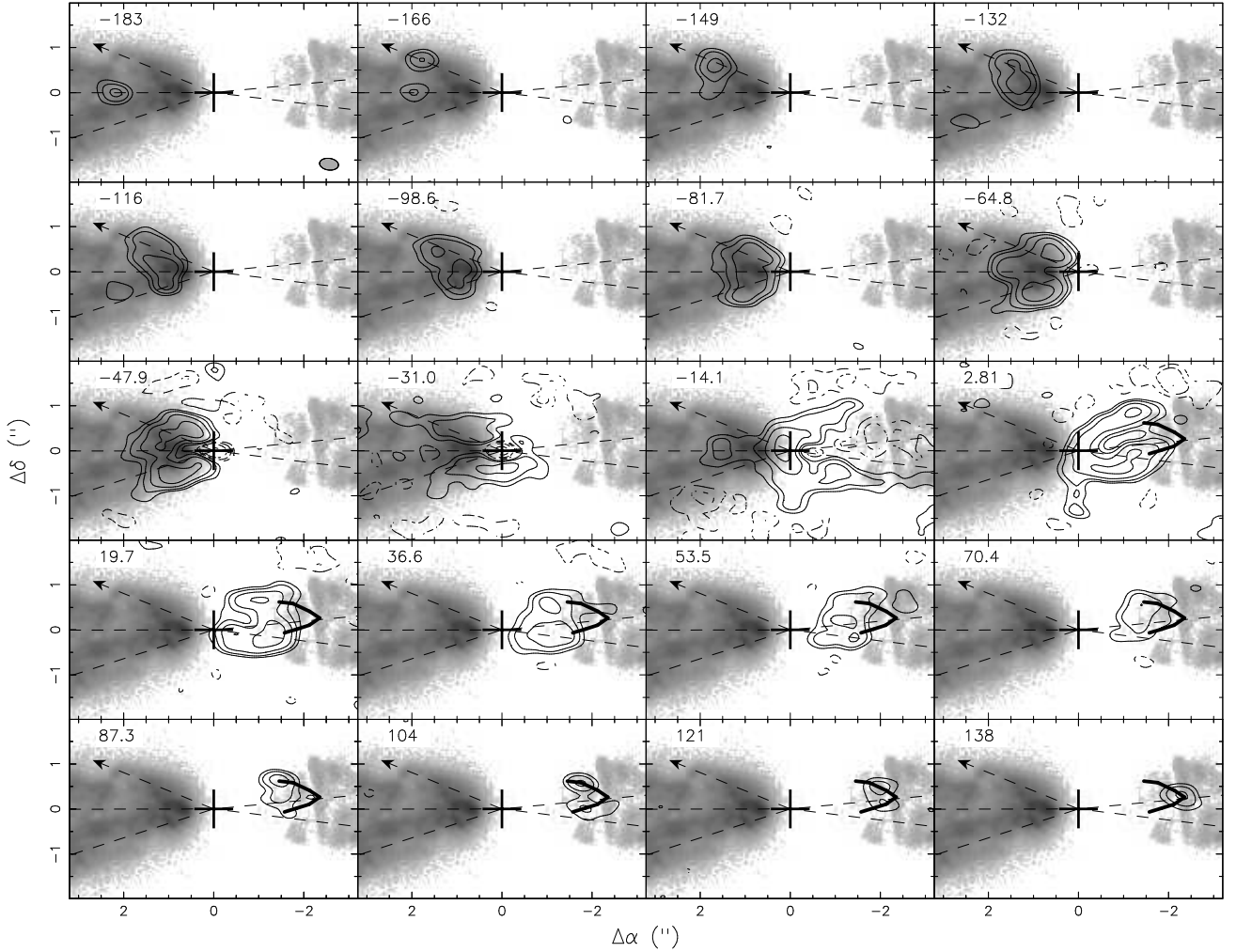

\centering
\putfig{0.8}{270}{f4.ps} 
\figcaption[]
{
CO channel maps on top of the H$\alpha$ image toward the center of CRL 618.
The cross and arrows have the
same meaning as those in Figure \ref{fig:Ha_CO}.  The angular resolution is
\arcsa{0}{42}$\times$\arcsa{0}{26}.  The bow-like structure is the same as
that in Figure \ref{fig:Ha_CO}.  The velocity in \vkm{} is indicated in the
top left corner in each panel. The contour levels are
$4\cdot2^{n-1}\sigma$, where $n=1,2,3..$, and $\sigma=12$ \mJyb{}.
Equivalent negative contours are plotted in dashed lines.
\label{fig:COchan}
} 
\end{figure}

\begin{figure} [!hbp]
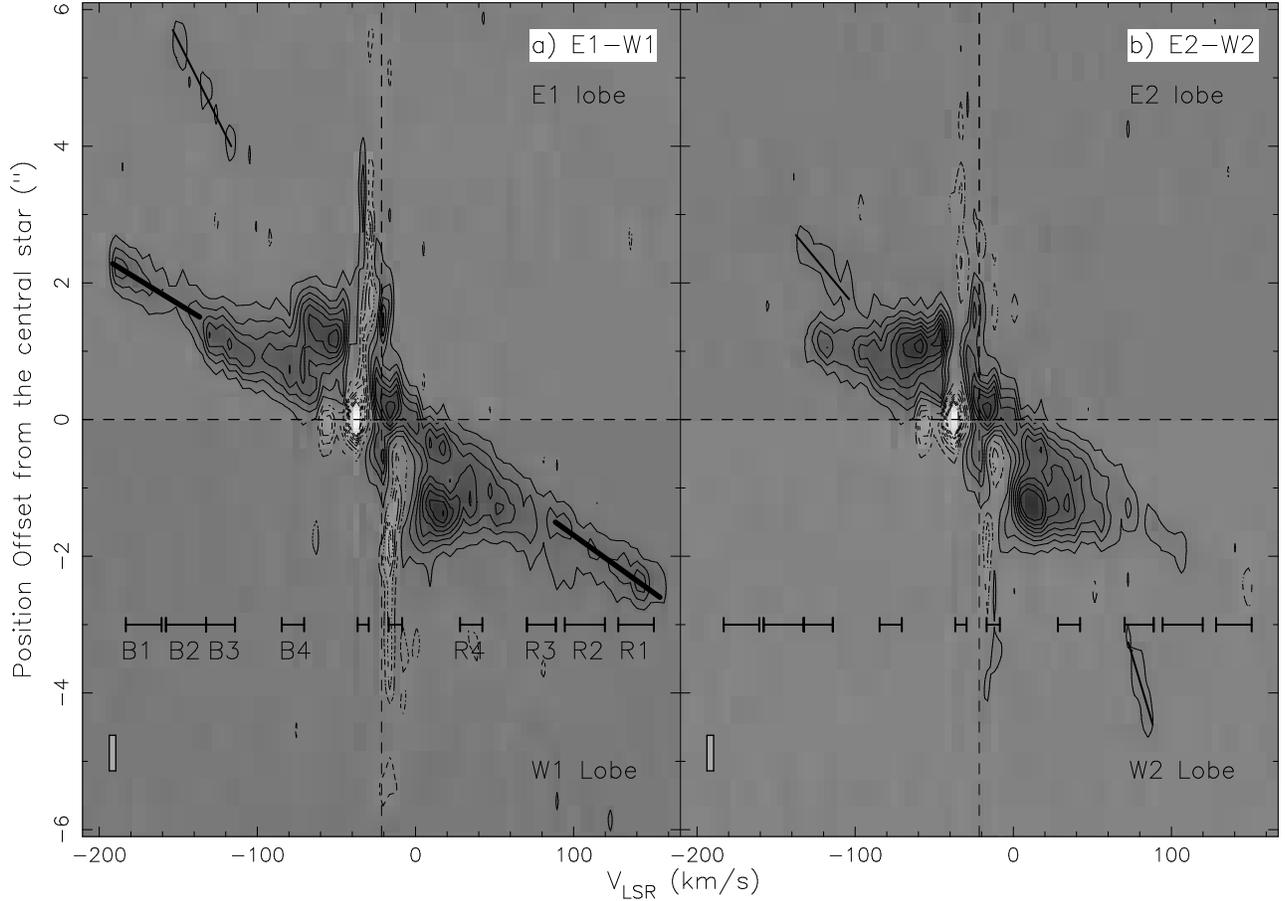

\centering
\putfig{0.7}{270}{f5.ps} 
\figcaption[]
{
Position-velocity (PV) diagrams of CO emission cut along the axes of the outflow
lobes.  (a) along the axes of the E1 and W1 lobes.  (b) along the axes of the
E2 and W2 lobes.  The vertical dashed line indicates the systemic velocity.
The linear PV structures at high velocity are indicated
with solid lines.  The horizontal bars indicate the velocity ranges used to
plot the high-velocity emission in Figure \ref{fig:Ha_CO}. The
contours start at 5 K with a step of 7 K. 
Equivalent negative contours are plotted in dashed lines.
The gray box in the lower left corner shows the velocity and spatial resolutions of the PV diagrams.
\label{fig:pvs_jet}
} 
\end{figure}

\begin{figure} [!hbp]
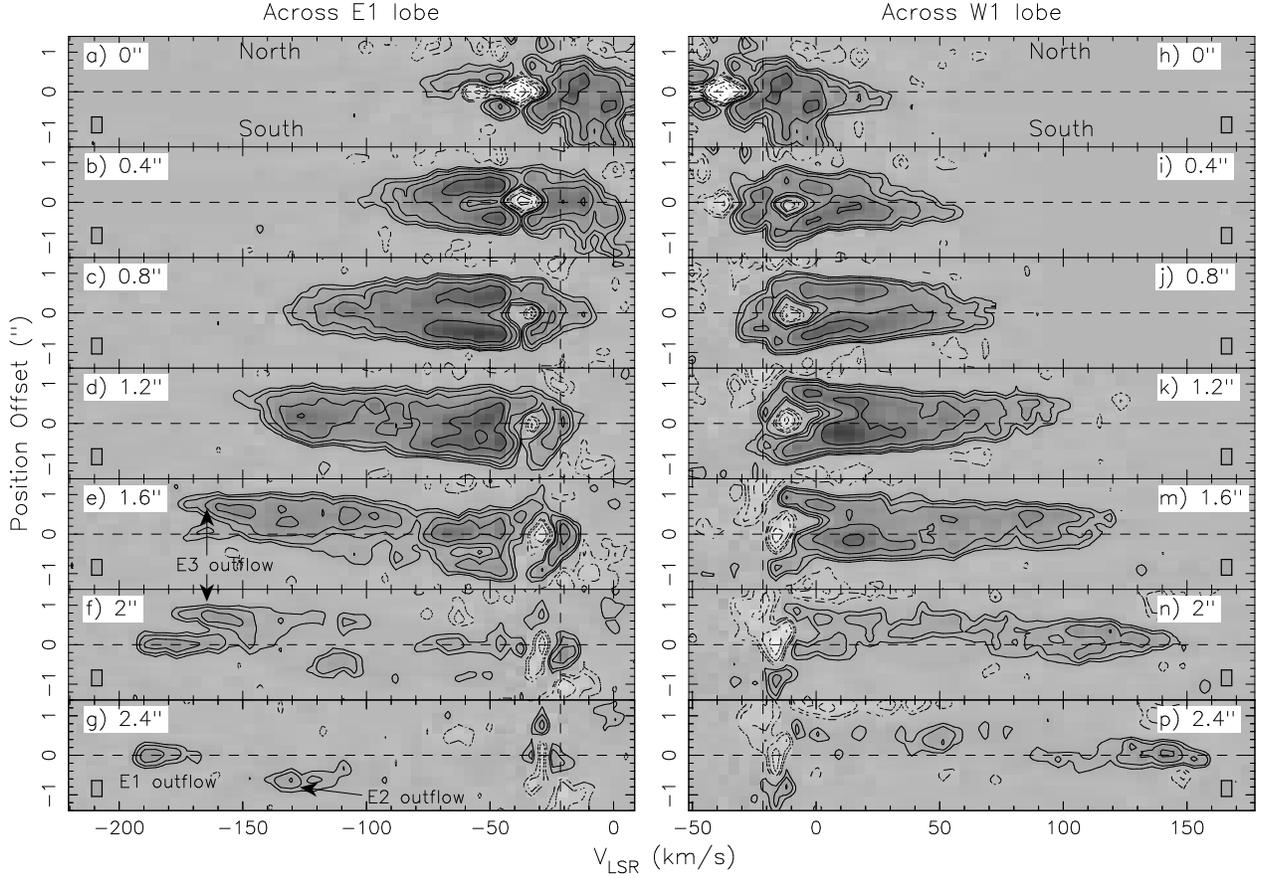

\centering
\putfig{0.65}{270}{f6.ps} 
\figcaption[]
{PV diagrams of CO emission cut across the outflow lobes at increasing distance from the central star. 
The distance is indicated in the top corner in each panel.
The vertical dashed line indicates the systemic velocity.
(Left) cut across the E1 lobe and (Right) cut across the W1 lobe.
The contour levels are
$2\cdot1.8^{n-1}\sigma$, where $n=1,2,3..$, and $\sigma=1.8$ K.
Equivalent negative contours are plotted in dashed lines.
The gray box in the lower corner shows the velocity and spatial resolutions of the PV diagrams.
\label{fig:pvs_perjet}
}
\end{figure}

\begin{figure} [!hbp]
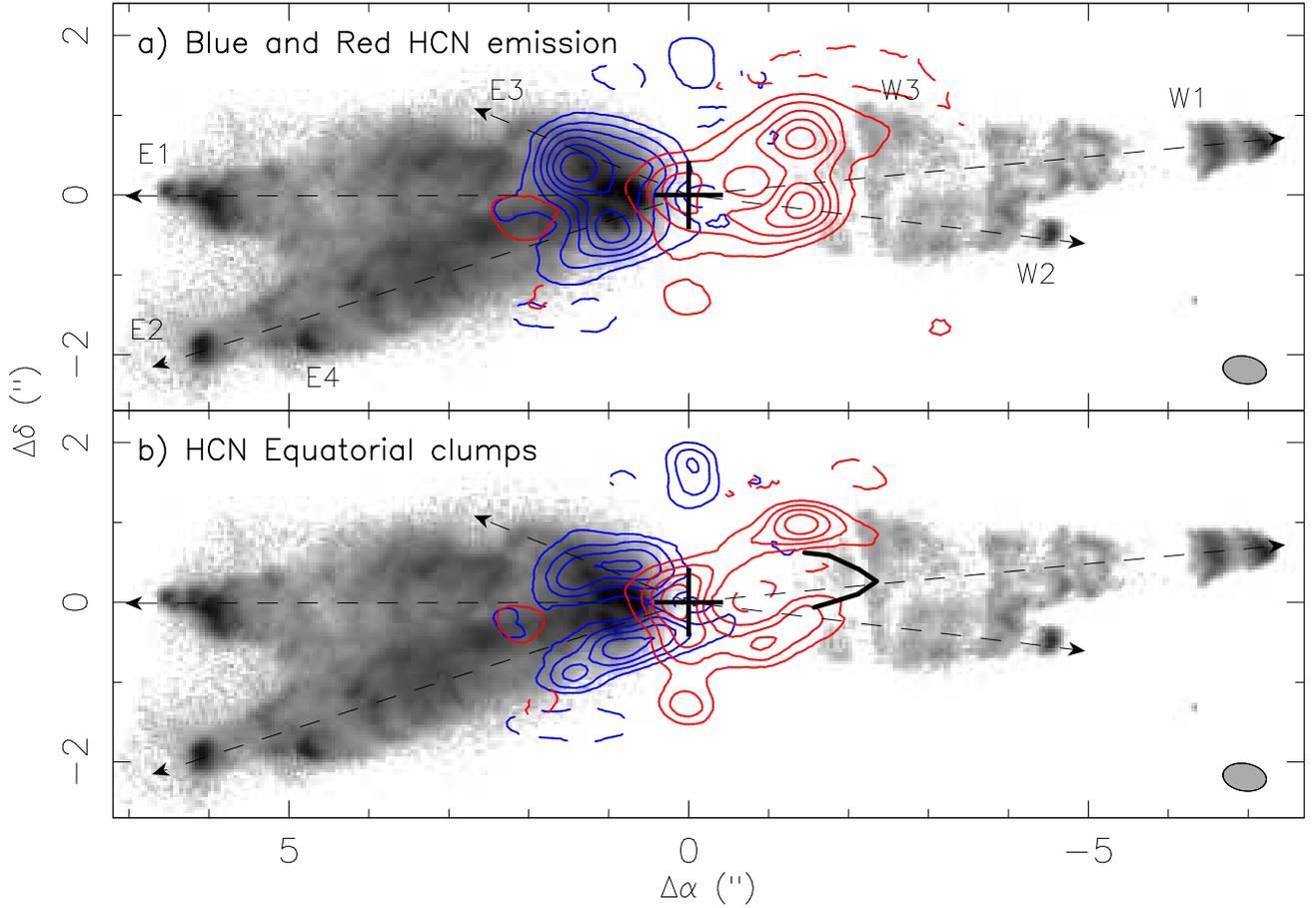

\centering
\putfig{1}{0}{f7.ps} 
\figcaption[]
{
Our SMA HCN maps plotted on top of the \Ha{} (gray-scaled) image.  (a) shows
the maps of the blueshifted (blue contours,
-138.9 to -22.5 \vkm{}) and
redshifted (red contours, -22.5 to 96.1 \vkm)
HCN emission.  Note that significant redshifted emission
around the central star is from the HCN $v_2=1$ J=4-3 line that traces the 
dense core \citep{Lee2013}.  The contours start at
5 \Jybk{} with a step of 15 \Jybk{}.  (b) shows the HCN emission at low
blueshifted (-48.2 to -22.5 \vkm) and redshifted (-19.3 to 1.3 \vkm)
velocity in order
to show the pair of molecular clumps in the equatorial plane.  The contours
start at 4 \Jybk{} with a step of 3.6 \Jybk{}.  The resolution of the HCN maps
is \arcsa{0}{54}$\times$\arcsa{0}{34}.  Equivalent negative contours are plotted in dashed lines.
\label{fig:eqoutflow}
}
\end{figure}

\begin{figure} [!hbp]
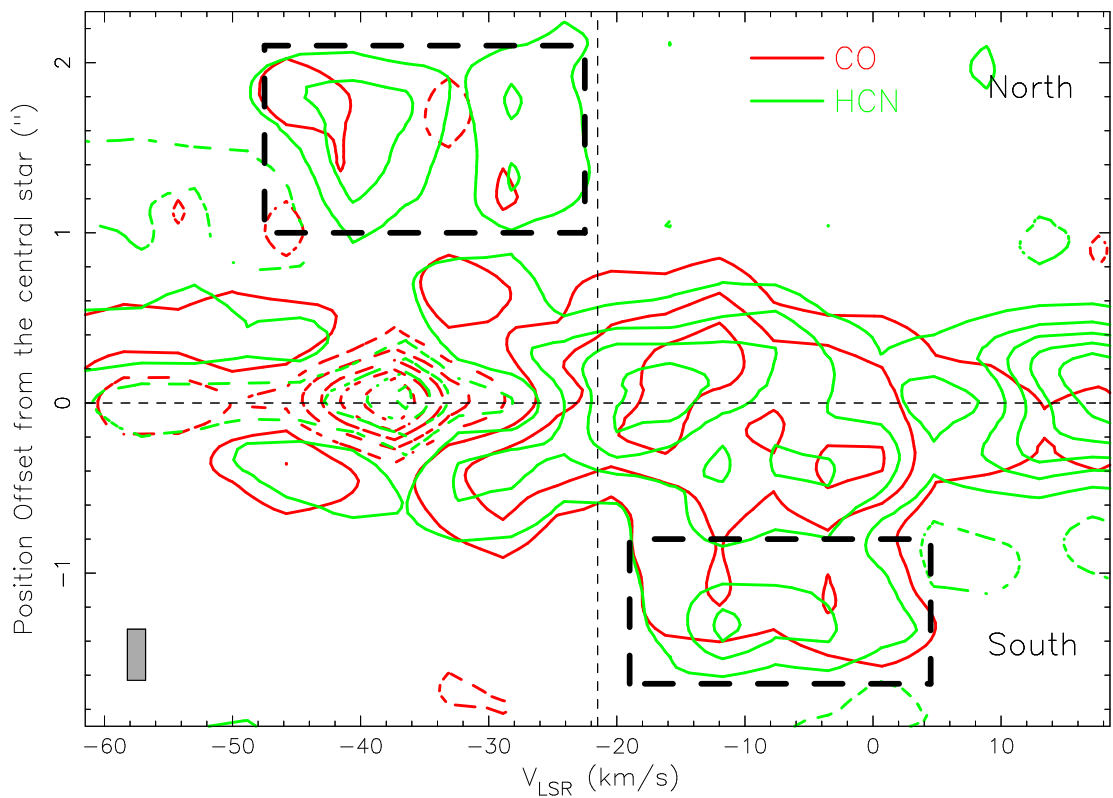

\centering
\putfig{0.6}{270}{f8.ps} 
\figcaption[]
{
PV diagrams of CO (red contours) and HCN (green contours)
emission cut along the equatorial plane through the central star position. 
The vertical dashed line indicates the systemic velocity.
The two dashed boxes mark the PV structures associated with the two
molecular clumps in the equatorial plane seen in Figure \ref{fig:eqoutflow}. 
The contours start at 8 K with a step of 16 K.  
Equivalent negative contours are plotted in dashed lines.
The gray box in the lower
left corner shows the velocity and spatial resolutions of the PV diagrams.
\label{fig:pvs_equatorial}
}
\end{figure}

\end{document}